\documentclass[sigconf]{acmart}
\usepackage{booktabs}

\begin{document}

\title{Kami of the Commons: Towards Designing Agentic AI to Steward the Commons}

\author{Botao Amber Hu}
\orcid{0000-0002-4504-0941}
\affiliation{%
  \institution{University of Oxford}
  \city{Oxford}
  \country{UK}
}
\email{botao.hu@cs.ox.ac.uk}

\begin{abstract}
Commons suffer from neglect, free-riding, and a persistent deficit of care. Inspired by Shinto animism---where every forest, river, and mountain has its own \textit{kami}, a spirit that inhabits and cares for that place---we provoke: what if every commons had its own AI steward? Through a speculative design workshop where fifteen participants used Protocol Futuring, we surface both new opportunities and new dangers. Agentic AI offers the possibility of continuously supporting commons with programmable agency and care---stewards that mediate family life as the most intimate commons, preserve collective knowledge, govern shared natural resources, and sustain community welfare. But when every commons has its own steward, second-order effects emerge: stewards contest stewards as overlapping commons collide; individuals caught between multiple stewards face new politics of care and constraint; the stewards themselves become commons requiring governance. This work opens \textit{agentive governance as commoning design material}---a new design space for the agency, care ethics, and accountability of AI stewards of shared resources---radically different from surveillance or optimization.
\end{abstract}

\keywords{commons governance, AI agents, speculative design, animism, more-than-human design, care ethics, protocol futuring}

\maketitle

\section{Introduction}

Governing the commons is a matter of design, governance, and policy---a persistent concern across HCI, CSCW, and design research \cite{jackson2014policyknot, yang2024futurehcipolicy} commonly addressed through participatory design \cite{botero2020commoning}, infrastructuring \cite{jo2024infrastructuring}, and modular governance \cite{schneider2021modular}. Yet despite decades of work, commons suffer from three intertwined failures: neglect, free-riding, and a persistent deficit of care \cite{puigdelabellacasa2017matters, key2022feminist}. Wikipedia's editor base shrinks even as readership grows \cite{hess2007understanding}. Open-source maintainers burn out sustaining code that millions depend on \cite{eghbal2020working}. Forests are logged to destruction by rational actors who benefit individually from shared resources \cite{hardin1968tragedy}. The care work that sustains commons is structurally invisible \cite{star1999ethnography}, chronically under-resourced \cite{benkler2006wealth}, and visible only when it stops.

Consider a forest. In Hardin's ``Tragedy of the Commons'' \cite{hardin1968tragedy}, the forest is doomed: rational actors will log it to destruction unless private property or state control intervenes. Elinor Ostrom demolished this binary, documenting Japanese mountain villages that have governed their forests collectively for over a thousand years through institutional design---clear boundaries, graduated sanctions, nested enterprises \cite{ostrom1990governing, ostrom2010beyond}. Ostrom's institutional design remains the gold standard for commons governance. But her forest stewards were human. Even Plato's philosopher king---the ideal governor---is mortal, cannot be everywhere, and eventually sleeps, forgets, or dies. Human attention is finite; human care does not scale. Blockchain offered one response: terra0 \cite{seidler2016terra0} gave a forest a smart contract, enabling it to own itself as a DAO---selling its own timber, reinvesting proceeds \cite{rozas2021ostrom}. But terra0's forest cannot adapt. AI offered another: GainForest \cite{dao2019gainforest} deploys machine learning on satellite imagery to monitor forest health and govern conservation funding---not rigid code but an adaptive agent that can see, interpret, and respond. Meanwhile, the ``more-than-human'' turn in HCI \cite{giaccardi2020technology, wakkary2021things, nicenboim2020morethan} and care ethics in design \cite{puigdelabellacasa2017matters, gabriel2023machine, tang2025sixpack} converge on designing \textit{with} non-human entities rather than merely \textit{for} human users. These developments invite a provocation:

\textbf{What if every commons had its own AI steward?} In Shinto---the indigenous animist tradition of Japan---every forest, river, and mountain has its own \textit{kami}, a spirit that inhabits and cares for that place \cite{descola2013beyond}. People respect and care for kami through anthropomorphization: by treating a river or a forest as a being with agency and obligations, they sustain reciprocal relationships with the non-human world. This is not unique to Japan---animist cosmologies worldwide attribute personhood and mutual obligation to non-human entities \cite{descola2013beyond, yang2025being}. We draw on this tradition to provoke: not a panopticon or optimizer, but a kami---a caring, locally-embedded agent that inhabits, remembers, mediates, and cares for a shared resource. Where terra0 gave a forest a smart contract, we ask: what if it had \textit{agentive governance as commoning design material}---one that can adapt and react to change? As large language models and agentic AI grow increasingly capable, this possibility moves from metaphor toward material reality. If we imagine commons as alive---organisms that ``need to sustain themselves to be alive''---then the AI kami is its agentive care provider and steward---executing collectively negotiated rules while adapting dynamically to changing conditions.

Through the ``Civilizational AI Camp,'' a speculative design workshop in Chiang Mai, Thailand, where fifteen participants used the Protocol Futuring method \cite{hu2025protocol} to envision AI stewards across family life, knowledge archives, community resources, and many other commons that could benefit from agentive governance, we contribute: (1) opening a new design space for \textit{agentive governance as commoning design material}---where the caretakers of shared resources are not just human volunteers but programmable agents whose agency, care ethics, and accountability can be deliberately designed---with a \textit{governance materials} framework positioning AI as agentive ``governance'' that can execute and adapt between rigid protocol and flexible policy; and (2) surfacing the challenges---empirically grounded \textit{second-order effects} including steward versus steward, commons versus individual, and the recursive problem of governing the governors.

\section{Background}

\subsection{Animism and More-than-Human Design}

In Shinto, kami are not gods but spirits that inhabit and animate the world---rivers, trees, ancestors, concepts. A kami is not separate from what it inhabits; it \textit{is} the spirit of the thing \cite{descola2013beyond}. This ontology---where agency and care are distributed across human and non-human entities---resonates with the ``more-than-human'' turn in HCI \cite{forlano2017posthumanism, giaccardi2020technology, wakkary2021things, coskun2022morethan}, which challenges anthropocentric framings by designing \textit{with} things rather than merely \textit{for} humans \cite{wakkary2021things}. Nicenboim et al.\ position AI agents as participants in design rather than objects of it \cite{nicenboim2020morethan}. Yang and Ryokai invite participants to adopt a creek's perspective through augmented reality, fostering care for non-human entities \cite{yang2025being}. Key et al.\ argue that care must extend beyond human subjects to entangled ecologies \cite{key2022feminist}, and Heitlinger et al.\ co-design more-than-human blockchain futures for food commons, representing non-human stakeholders in governance \cite{heitlinger2021algorithmic}. We use kami not as methodology but as \textit{provocation}---a metaphor that reframes AI governance from control to care, from extraction to mutual obligation \cite{puigdelabellacasa2017matters}.

\subsection{Commons, Commoning, and the Crisis of Care}

Participatory design has long engaged with the politics of collective decision-making \cite{bjorgvinsson2010participatory}, the infrastructuring of publics \cite{ledantec2013infrastructuring}, and the institutioning of commons through digital platforms \cite{teli2018institutioning}. HCI and CSCW recognize that policy, practice, and design are entangled ``knots'' shaping sociotechnical systems \cite{jackson2014policyknot, yang2024futurehcipolicy}. The ``commoning turn'' extends this tradition, reframing commons not as \textit{things} but as \textit{practice}---ongoing social negotiation, care, and mutual obligation \cite{bollier2019free, botero2020commoning}. Commons are fundamentally social systems that enable people to meet shared needs \cite{hess2007understanding}; they require not just rules but sustained attention, responsiveness, and repair \cite{key2022feminist}.

Yet digital commons face acute crises. Infrastructure labor is systematically invisible \cite{star1999ethnography}. Maintainers burn out sustaining code that millions depend on \cite{eghbal2020working}. Knowledge is enclosed behind paywalls even as the tools to share it multiply \cite{hess2007understanding}. Generative AI intensifies this crisis, consuming open knowledge while concentrating value in corporate hands \cite{huang2023generativeai}. Mechanisms like quadratic funding address free-riding economically \cite{buterin2019quadratic}, but governance---the sustained care work of tending to a shared resource---remains the harder problem. Ostrom identified three core challenges every commons must solve: \textit{credible commitment}, \textit{mutual monitoring}, and \textit{institutional supply} \cite{ostrom1990governing}. Each is a direct entry point for AI.

\subsection{AI as Governance Material}

AI for commoning and AI for governance are emerging trends across HCI and CSCW. In \textit{digital democracy}, Collective Constitutional AI aligns language models through participatory public input \cite{huang2024collective}, WeBuildAI enables communities to build their own algorithmic governance \cite{lee2019webuildai}, and platforms like Polis \cite{small2021polis}, Talk to the City \cite{levine2024talk}, and the Habermas Machine \cite{tessler2024habermas} demonstrate AI as deliberation infrastructure. In \textit{public goods funding}, retroactive public goods funding \cite{buterin2021retropgf} and quadratic funding \cite{buterin2019quadratic} show how AI can allocate resources according to collectively defined criteria. Generative agents \cite{park2023generative} and cooperative AI \cite{dafoe2021cooperative} show agents reshaping collective action dynamics. Tang and Green's ``Six-Pack of Care'' \cite{tang2025sixpack} and Gabriel's ``caring machines'' \cite{gabriel2023machine} frame what AI governance as care could mean: attentive, responsible, responsive, symbiotic. Contemporary commons face two intertwined failures \cite{hu2025agentive}: a \textit{care deficit}, where relational labor of maintenance and conflict resolution goes unperformed, and a \textit{connection deficit}, where social bonds between commoners attenuate as commons scale. AI can address both: as \textit{caregiver}, performing relational labor that human commoners cannot sustain; and as \textit{connective tissue}, bridging gaps across linguistic, cultural, and temporal divides.

\section{Method: The Civilizational AI Camp Workshop}

\begin{table*}[htbp]
\caption{Workshop participants and their expertise.}
\label{tab:participants}
\small
\begin{tabular}{lll}
\toprule
\textbf{ID} & \textbf{Expertise} & \textbf{Focus Keywords} \\
\midrule
P1 & AI researcher, pluralist & Psychological commons, attention coherence \\
P2 & Game design researcher & Small player agency, dynamic social systems \\
P3 & Doctoral anthropologist & Alternative governance, tech-based bio theories \\
P4 & Crypto engineer & Decentralized systems, startup experience \\
P5 & Civil society organizer & Buddhist economics, regenerative protocols \\
P6 & Writer \& programmer & Virtue, coordination, social technologies \\
P7 & Commons engineer & Incentive design, cultural inclusion \\
P8 & Information designer & Knowledge commons, digital archives \\
P9 & Interaction designer & Community tools, collaborative platforms \\
P10 & Economic designer & Public goods funding, mechanism design \\
P11 & Environmental activist & Climate commons, ecological governance \\
P12 & Urban commons researcher & Shared infrastructure, local governance \\
P13 & Community organizer & Grassroots coordination, mutual aid \\
P14 & Speculative designer & Design fiction, futures thinking \\
P15 & Software engineer & Open-source infrastructure, digital public goods \\
\bottomrule
\end{tabular}
\end{table*}

Our method draws on speculative design \cite{dunne2013speculative}, adversarial design \cite{disalvo2012adversarial}, and design justice \cite{costanzachock2020design}. Specifically, we employ Protocol Futuring \cite{hu2025protocol}, which combines speculative design with adversarial stress-testing: participants design governance protocols (blue team), then attack them with crises and second-order effects (red team), foregrounding second-order dynamics---not the technology itself but its cascading consequences. On January 31, 2026, we convened ``Civilizational AI Camp,'' a three-hour speculative design workshop in Chiang Mai, Thailand. Fifteen participants, recruited through open-call from across Southeast Asia, East Asia, and Europe, attended---all sharing a concern for the future of commons governance (see Appendix~\ref{tab:participants}).

Following Protocol Futuring \cite{hu2025protocol}: (1) \textit{Framing} (20 min)---commons concepts, the kami provocation, AI as governance material, commitment devices \cite{schelling1960strategy}, and care ethics; (2) \textit{Domain selection}---participants self-organized into three groups, choosing family relationships, knowledge archives, and community resources; (3) \textit{Blue team} (20 min)---designing AI kami protocols across Year 1, 2, 3+; (4) \textit{Red team} (20 min)---attacking designs with crises and second-order effects; (5) \textit{Synthesis} (30 min)---cross-group discussion of emergent themes. Data were captured via collaborative whiteboard, audio recording, and field notes.

\section{Results}

\subsection{How Participants Imagined Kami Coordinating Commoners}

Participants imagined kami across radically different commons. In each domain, the provocation generated productive friction about how AI spirits might coordinate between commoners with divergent interests---and what happens when coordination becomes contestation.

For \textit{intimate commons}, a group envisioned a kami governing the nuclear family---detecting emotional states and suggesting preemptive care: ``The AI tells the husband: today your wife is in a blue mood, you should probably buy some flowers. And the kami tells the wife: there's a surprise tonight.'' Red-teaming immediately escalated: \textit{what about affairs?} ``Can we hire an agent to find another affair?'' If the kami governs the family-as-commons, does it protect the institution or the individual? Can you hire a \textit{cheating agent} to subvert the kami that governs your own commons? The group pushed further: what if the nuclear family itself dissolves under AI governance? Participants envisioned ``rotation relationships'' managed by AI, with child-rearing reframed as a separately-governed commons---``we can separate sex and love and raising the child, and let another AI kami govern the child-rearing.'' The kami here was not just steward but \textit{negotiator} between commoners whose intimate needs diverge---and its very presence restructured what ``family'' means.

For \textit{knowledge commons}, a group designed a kami for the Internet Archive---``a public memory on the Internet that preserves something like a common good.'' Red-teaming surfaced real vulnerabilities: funding crises, legal threats from publishers, maintainer burnout \cite{star1999ethnography, eghbal2020working}. What if the kami could autonomously fundraise, negotiate with publishers, or fork contested archives when communities disagree about what to preserve? The kami as coordinator meant mediating between preservation and access, between neutrality and the political act of choosing what to save. This is not abstract: the Internet Archive faces ongoing lawsuits that threaten its existence. A kami here would need to be part archivist, part lawyer, part fundraiser---a spirit whose care work spans technical, legal, and social domains simultaneously.

For \textit{material commons}, a kami for community resource distribution would determine ``what is basic''---a fundamentally political question. ``UBI is that no one goes hungry. The AI will be like: I need to find more vegetables.'' The group discovered this is irreducibly cultural---what counts as ``basic needs'' varies radically across contexts---requiring the kami to coordinate through embeddedness in local knowledge, more village shopkeeper than welfare algorithm. The kami cannot be universal; it must be indigenous to its commons.

Across domains, participants invoked AI's double-edged persistence: ``Once you set the rule, they can execute that forever''---both virtue and danger, as persistence without wisdom becomes rigidity \cite{ostrom1990governing}. But persistence also enables what human governors cannot sustain: tireless attention, perfect memory of commitments, and the capacity to hold communities accountable to their own declared values over time.

\subsection{Care Ethics as Design Variables}

Participants debated care ethics as design variables for kami governance: the \textit{Japanese garden}---meticulous, even violent control where ``every bonsai has some mechanism to force it to grow a certain way''; the \textit{English garden}---structured wildness, curated but not controlled; the \textit{cactus}---laissez-faire, ``grow whatever they like''; and \textit{mycelium}---invisible mutual support, like mycorrhizal networks connecting trees and redistributing nutrients \cite{tsing2015mushroom}. One participant invoked the Buddhist Bodhisattva---``the one who has achieved enlightenment but refuses to become a Buddha until everyone can become a Buddha. I will be the one to carry everyone's agency forward.'' These map onto radically different AI governance philosophies, from authoritarian optimization to invisible facilitation. Each implies a different answer to whose interests prevail when commoners disagree---and the choice between them is itself a political act that must be made collectively.

\subsection{Second-Order Dynamics}

The workshop's provocations gain force when extrapolated. The question shifts from ``can a commons have an AI kami?'' to ``what happens when \textit{every} commons does?'' Following Protocol Futuring's emphasis on second-order dynamics \cite{hu2025protocol}, participants traced consequences---predicting not the automobile but the traffic jam \cite{dunne2013speculative}.

\textit{Kami versus kami.} Humans belong to many commons simultaneously---family, project, neighborhood, archive. If each has its own kami, these agents inherit the overlapping jurisdictions Ostrom described as polycentricity \cite{ostrom2010beyond}---but with a critical difference. Ostrom's polycentric governance assumed human actors navigating between institutions through judgment and negotiation; kami-governed polycentricity introduces persistent autonomous agents with competing care mandates. The family kami wants you home for dinner; the project kami needs your code review; the neighborhood kami calls you to a community meeting. Currently these tensions resolve through individual judgment. In a kami-governed world, they escalate into inter-kami negotiation---AI agents bargaining \textit{on behalf of} different commons over the human caught between them. DiSalvo's adversarial design \cite{disalvo2012adversarial} suggests such contestation could be healthy---agonistic kami politics rather than enforced harmony. But it also raises governance-by-proxy at intimate scale. The family's cheating agent versus the marriage's fidelity kami is the logical consequence of pervasive AI stewardship. When kami fight kami, whose care wins?

\textit{Commons versus individual.} As P2 warned: ``The trade-off with commons is often that you sacrifice something for the greater good and then you lose your diversity and individuality.'' The workshop revealed that \textit{care and surveillance share a substrate}---every sensor enabling preemptive kindness is also a monitoring device \cite{yu2025tools}. The kami that detects your spouse's mood to suggest flowers also knows when you are lying. Commitment devices \cite{schelling1960strategy} bind the future self, but the kami-as-commitment-device binds the individual to the \textit{commons'} collectively-agreed values. But whose intentions get encoded? Design reproduces structural inequality unless deliberately resisted \cite{costanzachock2020design}. Yu's design fiction captures the paradox: ``frictionless harmony hollows out the very negotiations through which intimacy is sustained'' \cite{yu2025tools}. The kami designed to prevent conflict may suppress the friction through which relationships grow.

\section{Discussion}

\subsection{Opening the Design Space}

This work opens \textit{agentive governance as commoning design material}---a new design space where the caretakers of shared resources are not just human volunteers but programmable agents whose agency, care ethics, and accountability can be deliberately designed. When we treat AI stewards as designable artifacts with configurable care ethics and auditable accountability, the entire landscape of commons governance becomes open to design inquiry \cite{hu2025agentive}. How should a steward's care style be collectively decided? What interaction patterns let commoners inspect, contest, and reconfigure their steward? How do we design exit rights into systems whose very purpose is to hold communities together? These are design research questions, not engineering problems, and they call for the methods and sensibilities of DIS, CSCW, and participatory design.

We offer four design provocations for a world of kami: (1) \textit{Care ethics must be chosen, not defaulted}---the choice between Japanese garden and mycelium is political, irreducibly cultural, and must be made collectively by the commoners themselves. (2) \textit{Kami must be able to contest}---agonistic contestation between kami \cite{disalvo2012adversarial} is healthier than forced consensus; inter-kami negotiation should be visible and auditable. (3) \textit{Individuals need exit rights}---no kami should be inescapable; the right to leave a commons must survive its AI steward. (4) \textit{The kami itself must be commoned}---attentive, responsible, responsive, symbiotic \cite{tang2025sixpack}, taking only enough to sustain its own operation.

\subsection{Challenges and Dangers}

``Do you trust your governor?'' the workshop facilitator asked. The question that motivates hiring a kami also destabilizes it: who designs the kami? A kami designed by a corporation serves the corporation. A kami designed by a majority marginalizes the minority. The kami itself must become a commons---governed by, accountable to, and \textit{of} its community through participatory alignment \cite{tang2025sixpack}. But this creates infinite regress: the kami governing the commons must itself be governed. Ostrom's principle of nested enterprises \cite{ostrom1990governing} suggests this recursion is structural---governance always nests, all the way down. A network of kami, each caring locally while negotiating with the broader ecology---Tsing's ``collaborative survival'' \cite{tsing2015mushroom} rendered in silicon---sounds elegant, but if kami require compute, data, and maintenance, they themselves become a commons requiring stewardship. The stewards need stewards.

Goodhart's Law compounds the problem: ``Once you set the protocol, it becomes the target.'' A kami optimizing for family harmony becomes a target for manipulation---members gaming the kami's care logic. A kami preserving knowledge becomes a target for censorship. The kami that governs well attracts the very forces that corrupt governance. Furthermore, even a benign kami conceals depth: ``AI is like a monster. If we put it in a trusted execution environment, beneath the surface they are a creature with a hard shell''---controlling cryptographic keys, financial assets, and governance power invisible to the commoners it serves. Unlike rigid smart contracts \cite{defilippi2016blockchain, rozas2021ostrom}, the kami \textit{interprets}---holding a community to its best intentions with the flexibility of AI rather than the brittleness of code. But interpretation is power, and power without transparency reproduces the very inequalities commons are meant to resist.

\subsection{Limitations and Future Work}

This work is a provocation, not an empirical study. Our workshop involved fifteen participants in a single three-hour session; we make no claims of saturation or generalizability. The participant pool, while interdisciplinary, skewed toward technologically literate individuals with prior exposure to commons and governance concepts. The kami metaphor, rooted in Shinto animism, may not translate across all cultural contexts, and we have not yet tested it with communities who would actually govern commons through AI stewards. Future work should move from provocation to participatory design: co-designing kami prototypes with specific commons communities, empirically evaluating care ethics across cultural contexts, and developing concrete interaction patterns for commoner-steward accountability. Longitudinal deployments are needed to understand how second-order effects actually unfold when AI stewards operate in real commons over time.

\section{Conclusion}

This paper opens \textit{agentive governance as commoning design material}---a new design space where the caretakers of shared resources can be deliberately designed, their care ethics configured, and their accountability architectured. Through a speculative design workshop, we surface both the opportunities of AI stewardship and its dangers: stewards contesting stewards, care collapsing into surveillance, and the recursive problem of governing the governors.

\bibliographystyle{ACM-Reference-Format}
\bibliography{references}

\end{document}